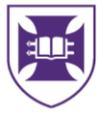
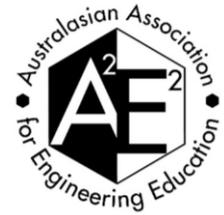

# Bridging Qualitative Rubrics and AI: A Binary Question Framework for Criterion-Referenced Grading in Engineering

Lili Chen, Winn Wing-Yiu Chow, Stella Peng, Bencheng Fan and Sachitha Bandara
*The University of Melbourne*
lili.chen1@unimelb.edu.au

## CONTEXT

Recent advancements in generative artificial intelligence (GenAI) have increasingly influenced educational practices, particularly in grading and feedback of written assessments. Prior research has shown that GenAI offers promising avenues to enhance assessment practices. While GenAI has increasingly advanced in solving mathematical problems in engineering education, its potential to assist with grading for mathematical assessments remains underexplored.

## PURPOSE OR GOAL

This study investigates how GenAI can be integrated with a criterion-referenced grading framework to improve the efficiency and quality of grading for mathematical assessments in engineering. It specifically explores the challenges demonstrators face with manual, model solution-based grading and how a GenAI-supported system can be designed to reliably identify student errors, provide high-quality feedback, and support human graders. The research also examines human graders' perceptions of the effectiveness of this GenAI-assisted approach.

## APPROACH OR METHODOLOGY/METHODS

A mixed-methods approach was applied to mathematical exercises in an electrical engineering subject. Pre-testing qualitative reflections were gathered from two of the researchers, drawing on their experience as subject demonstrators, to identify challenges with the existing grading method. Based on these insights, a novel criterion-referenced grading method was developed by converting qualitative rubrics into "Yes/No" questions. This method was then evaluated by the same two researchers alongside a university-provided GenAI tool. Post-testing reflections were also gathered from the two researchers regarding the grading method's effectiveness. The approach was applied to 50 student submissions across 15 questions.

## ACTUAL OR ANTICIPATED OUTCOMES

The study found that GenAI achieved an overall grading accuracy of 92.5%, comparable to two experienced human graders. The two researchers, who also served as subject demonstrators, perceived the GenAI as a helpful second reviewer that improved accuracy by catching small errors and provided more complete feedback than they could manually. A central outcome was the significant enhancement of formative feedback. However, they noted the GenAI tool is not yet reliable enough for autonomous use, especially with unconventional solutions.

## CONCLUSIONS/RECOMMENDATIONS/SUMMARY

This study demonstrates that GenAI, when paired with a structured, criterion-referenced framework using binary questions, can grade engineering mathematical assessments with an accuracy comparable to human experts. Its primary contribution is a novel methodological approach that embeds the generation of high-quality, scalable formative feedback directly into the assessment workflow. Future work should investigate student perceptions of GenAI grading and feedback.

# Introduction

The integration of generative artificial intelligence (GenAI) into education offers promising avenues to enhance assessment practices. Research highlights several key benefits of GenAI-assisted grading, including its ability to process large volumes of student work, apply marking criteria consistently, provide personalised feedback, and minimise common human errors such as fatigue or oversight (Li et al., 2024). These advantages are particularly valuable in large-scale educational settings, such as undergraduate programs with high enrolments and demanding assessment loads, where academics often face challenges in delivering timely, equitable, and high-quality grading with formative feedback.

Engineering assessments typically include problem sets, written tests, and exams that emphasise students' mathematical reasoning and problem-solving. Grading such work requires evaluating complex reasoning, multi-step solutions, and symbolic notation. In this context, GenAI's problem-solving capability, scalability, and consistency could potentially reduce grading burdens and enhance feedback quality. However, while GenAI has been widely applied to grading written assignments and multiple-choice tests, its use for assessing mathematical work, a major form of assessment in engineering education, remains largely unexplored (Pepin et al., 2021).

Recent developments in GenAI have demonstrated impressive capabilities in solving complex mathematical problems. Newly released models, such as OpenAI's ChatGPT o3 and Google's Gemini Pro 2.5, exhibit advanced proficiency in symbolic reasoning, multi-step calculations, and articulating mathematical solutions in natural language. For instance, on the AIME 2025 benchmark, which assesses high-school-level mathematical skills, Gemini Pro 2.5 scored 88% and ChatGPT o3 achieved 88.9%, rivalling or surpassing human performance (Google, 2025). These results underscore GenAI's growing ability to tackle non-trivial mathematical tasks and its potential to support assessment processes requiring both procedural accuracy and conceptual clarity.

Engineering mathematics assessments are often graded using model solutions, where student work is compared against a single exemplar answer annotated with a marking scheme. However, this method often lacks explicitly defined performance criteria or detailed qualitative rubrics, leading to inconsistencies and less effective formative feedback (Morphett et al., 2019). By contrast, criterion-referenced grading, commonly used for written assignments, evaluates work against a predefined rubric that describes varying levels of performance (Lok et al., 2016). By specifying expectations and identifying strengths and areas for improvement, this method prioritizes the demonstration of conceptual understanding and application, while enhancing grading consistency and strengthening formative feedback. In a study on undergraduate statistics assessments, criterion-referenced grading achieved comparable inter- and intra-rater reliability to traditional model-solution marking (Morphett et al., 2019). Despite these benefits, this method remains underexplored and underutilized in engineering mathematics assessment.

Therefore, this study aims to evaluate GenAI-assisted criterion-referenced grading in engineering mathematics assessment. Specifically, two research questions (RQs) were explored:

- **RQ1:** What challenges do academics encounter in manually marking engineering mathematics assessments using a model solution-based grading approach?
- **RQ2**: How can GenAI be effectively designed or configured to support criterion-referenced grading in engineering mathematics assessments?

This study consisted of two main phases. The first phase investigated the challenges in the current model solution-based grading process of mathematics assessments in an electrical engineering subject at an Australian university. The second phase designed and explored the use of GenAI combined with criterion-referenced grading for the subject and evaluated its performance using both quantitative and qualitative data.

## Literature Review

### Engineering Mathematical Assessment

Enhancing engineering students' mathematical skills has long been a central focus of discussion and research in engineering education (Gray et al., 2024). A key driver of this focus is the observed decline in the fundamental mathematical competencies of incoming engineering students. Combined with the logistical challenges of teaching and assessing large cohorts, this has prompted the development of innovative pedagogical and assessment strategies. Recent literature emphasizes a shift from traditional summative examinations to more frequent, formative assessments that drive continuous learning, provide timely feedback, and strengthen the connection between mathematical concepts and engineering practice (Ní Fhloinn & Carr, 2017).

Various formative assessment strategies have been proposed, including in-class exercises that allow observation of student work and provide immediate feedback; low-stakes homework assessments designed to encourage consistent effort, often using partial marking strategies to reduce instructor workload; and peer-to-peer teaching methods(e.g., "jigsaw") which promote cooperative learning (Ní Fhloinn & Carr, 2017). A common thread across these approaches is their emphasis on delivering formative feedback, whether from instructors or peers, that helps students identify mistakes, refine reasoning, and develop problem-solving strategies. However, delivering high-quality, personalized feedback is time-intensive, especially in large cohorts, highlighting the need for scalable solutions that enhance learning while reducing grading burdens.

Before the emergence of GenAI, computer-assisted assessment provided a degree of automation for large-scale mathematics formative assessments by using extensive question banks, often containing thousands of items organized by topic and difficulty, and simple answer-entry interfaces (Chirwa, 2008; Davis et al., 2005). Students typically submitted numerical answers and received immediate outcomes. Although studies reported a positive correlation between the use of these systems and student performance, they were criticized for perceived unfairness and limited support for learning. Specifically, they evaluated only the final answer, provided no partial credit for correct reasoning or minor errors, and offered minimal formative feedback to guide student improvement.

### GenAI-Assisted Assessment

With the rise of GenAI, the landscape of computer-assisted assessment has evolved dramatically. Unlike earlier systems that only evaluated final answers, modern GenAI models such as ChatGPT can interpret full student responses and generate personalized, formative feedback. This has transformative potential for engineering education, especially in mathematics, where stepwise reasoning is key. In a study by Puthumanaillam et al. (2025), GPT-4 achieved a B-grade (82.24%) on a control systems course with minimal-effort prompting, close to the class average of 84.99%. Other recent studies exploring large language models (LLMs) for grading and feedback in math assessments report both promise and challenges. GenAI has shown improved grading performance, especially after fine-tuning. For instance, Nilsson and Tuvstedt (2023) found that GPT-4 reached 75% accuracy in grading introductory programming. More recently, Zhao et al. (2025) fine-tuned LLMs for proof-by-induction problems; the top-performing Llemma-based models exceeded 90% rubric-based accuracy, outperforming most human graders (86.6%).

However, in real submissions, models sometimes failed unpredictably and operated as "black boxes" with limited transparency. For example, Kortemeyer et al. (2024) found that while AI could reliably identify passing thermodynamics exams, with a true positive rate above 94%, it failed to recognize more than half of the students who had passed, resulting in an overall accuracy of just 62%. Moreover, while AI-generated feedback was rubric-aligned, students often found it too generic and insufficiently specific to help them diagnose errors in their reasoning. Additional issues in handwritten conversion complicate AI grading in engineering. A significant technical barrier remains in converting handwritten student mathematical answers into machine-readable formats (Kortemeyer et al., 2024). Furthermore, the reliability of AI grading is highly sensitive to the granularity of the rubric. When grading multipart problems with numerous fine-grained rubric items, the models struggled to maintain consistency across multiple criteria. Conversely, approaches that relied on matching full solutions to ideal responses produced overly simplistic evaluations, lacking the nuance required to capture partial correctness or alternative valid methods. These limitations highlight the complexity of aligning AI assessments with real-world educational standards.

**Assessment Grading Approaches**

Assessment grading typically follows one of two main approaches: criterion-referenced or norm-referenced. Criterion-referenced grading evaluates student work against explicit performance criteria derived from intended learning outcomes. In contrast, norm-referenced grading assesses student performance relative to peers, often using a predetermined distribution of marks (Lok et al., 2016). While norm-referenced grading can help differentiate between high- and low-performing students, criterion-referenced grading offers clearer insight into individual progress toward learning goals, making it especially effective for formative assessment.

In engineering education, assessments often rely on model solution grading, where student responses are compared against one or more ideal answers using a marking scheme. This approach is common because engineering problems frequently have clear, well-defined solutions, making it relatively straightforward to create standardised answers that support efficient and seemingly objective grading. However, this method presents several issues. It tends to focus on procedural accuracy rather than conceptual understanding, limiting its effectiveness in assessing higher-order thinking or problem-solving. Moreover, model solutions often fail to account for valid alternative approaches, potentially penalising students who use unconventional but sound reasoning. The approach can also lead to inconsistent allocation of partial credit, particularly when student responses deviate from the model in non-trivial ways (Holmes & Smith, 2003). Additionally, feedback in this model is typically minimal, often limited to brief comments or annotations on student submissions, and lacks the structure and depth required to effectively support learning (Morphett et al., 2019). For these reasons, criterion-referenced grading, implemented with a clearly defined rubric, is considered a more suitable alternative to conventional model solution grading. It promotes greater consistency and accuracy in marking while providing more constructive formative feedback to support student learning.

# Methodology

This study employed a mixed-methods approach to evaluate the integration of GenAI with criterion-referenced grading in the context of weekly pre-laboratory engineering mathematics exercises in an undergraduate electrical engineering subject at an urban Australian university. The study consisted of two phases: (1) investigation of current assessment practices and associated challenges, and (2) evaluation of the use of GenAI to support criterion-referenced grading. The study was conducted under research ethics protocol 29248 secured by the Faculty's Teaching and Learning Laboratory.

Pre-laboratory mathematical assessments were selected as the focus of this study for several key reasons. These assessments played a critical role by ensuring that students possess the foundational knowledge necessary to engage meaningfully with practical laboratory work. Moreover, they were inherently formative. Their primary purpose was not merely to assign grades, but to provide timely feedback, helping students identify and address misconceptions before applying concepts in a hands-on context. As such, pre-lab mathematical assessments presented an ideal context for evaluating the potential of AI to support formative assessment grading.

**Phase 1: Exploring Current Practice** – A qualitative approach was employed in this phase, with primary data recorded by two researchers, who served as subject demonstrators. They reflected on their existing grading practices using a set of predefined questions, such as: "*What is your perception of the consistency of marking among different demonstrators for these exercises?*" These reflections offered rich, in-depth insights into the marking process, the challenges experienced, and the extent to which criterion-referenced evaluation is currently applied. Thematic analysis was used to identify common themes and patterns emerging from the researchers' reflections (Braun & Clarke, 2006).

**Phase 2: Evaluating UniAI (Claude 3.5 Sonnet) as a Tool for Criterion-referenced Grading** – This phase focused on the design, implementation, and evaluation of UniAI (pseudonym), a university-provided generative AI tool powered by the Claude 3.5 Sonnet model. The aim was to explore its potential to support criterion-referenced grading of pre-lab mathematical exercises using a mixed-methods approach. Fifteen questions were selected from five different pre-lab exercises in a core third-year Bachelor of Science subject, and first-year Master of Engineering subject covering linear systems, signal processing, and transform methods. The questions fell into four categories: (1) Numerical answer type (NAT) questions (40%) – requiring precise numerical values; (2) Descriptive reasoning questions (13%) – requiring concept explanations or method justification; (3) Short answer questions (13%) – assessing conceptual understanding through brief responses; and (4) Proof questions (33%) – testing logical reasoning and mathematical rigor. This distribution closely matched the subject's final exam, making it representative for evaluating AI's grading ability. A representative example of proof questions selected is: "*Which of the following discrete-time systems are linear in the input-output trajectories (v, y)? Justify your answer with either a proof or a counterexample*". Ten student submissions per pre-lab were sampled, yielding a dataset of 50 responses to evaluate AI's effectiveness in interpreting and grading answers using a rubric focused on conceptual understanding, reasoning, and problem-solving communication. A sample grading rubric is provided in Table 1.

**Table 1: Grading Rubric for a Pre-Lab Question on Linearity**

| Learning outcome | | Apply the principles of additivity and homogeneity to formally prove the linearity of a given system. |
|---|---|---|
| **Level** | **Marks** | **Description** |
| Level 1: Beginning | 0-2 | The work may fail to correctly identify the system as linear or provide correct proofs for additivity or homogeneity. Shows a fundamental lack of understanding. |
| Level 2: Developing | 3-4 | Correctly identifies the system, but the proofs for additivity or homogeneity are incomplete or contain errors. The work may have incorrect notation or gaps in reasoning. |
| Level 3: Accomplished | 5 | Correctly identifies the system as linear, provides clear and correct proofs for both additivity and homogeneity, and uses proper notation throughout. |

In addition to assigning marks, AI was also prompted to generate explanatory feedback, including its grading rationale and identification of student errors. To assess grading accuracy and consistency, two researchers independently graded the same set of student responses using the same rubric, providing a benchmark for comparison with AI's performance.

The quantitative results from the evaluation focused on grading consistency, measured by inter-rater reliability, and grading accuracy, assessed by comparing AI-generated scores with those of the researchers. In parallel, the same two researchers reflected on the usefulness of the AI-generated grading and the perceived clarity and pedagogical value of the explanations provided by AI. All student data were fully anonymized before analysis.

## Results

**Pre-testing reflections.** Three key themes emerged around the challenges of using model solution grading in the current marking process. (1) *Inconsistent Marking and Inadequate Guidance*: A primary concern for the two researchers, who also served as demonstrators, was the struggle to maintain marking consistency, especially when faced with student work that was ambiguous or unconventional. The problem was exacerbated by marking schemes that were often too rigid, sometimes based on "one particular solution rather than rubrics or criteria." This lack of guidance for different yet valid methods forced the researchers to "trust our instincts" and rely on their own professional judgment, leading to variations in how marks were awarded. (2) *Time Constraints and Superficial Feedback*: The researchers felt that severe time constraints made it hard to offer detailed comments. As a result, feedback often focused on simply identifying students' errors rather than explaining "why it is crucial to include that reasoning to get the full marks" or offering broader, more constructive guidance for future improvement. (3) *Illegible Handwriting and Marking Barriers*: The two researchers reported that poor or illegible handwriting could be a "significant barrier to accurate marking." Deciphering messy handwriting added an "extra layer of cognitive load" to the assessment process, making it more difficult and time-consuming to fairly evaluate the student's understanding.

**Our criterion-referenced grading design.** Drawing on the pre-reflection findings and literature, we developed a criterion-referenced grading approach using AI support. Through iterative testing and insights from Kortemeyer et al. (2024), we refined this method for the pre-lab exercises. The key steps are outlined below:

- We first converted both handwritten and typed student submissions into LaTeX using OCR software, followed by manual verification. Since LaTeX is a widely used typesetting language on which many AI models are trained, this step helped ensure consistent and reliable parsing of mathematical content (Puthumanaillam et al., 2025).
- To align with the rubric and ensure consistent grading, each question was mapped to 3–6 targeted "Yes/No" grading questions with fixed mark values. These grading questions were designed around common checkpoints typically used during marking and could be completed by an experienced marker in just a few minutes. Figure 1 shows the five questions used for the rubric in Table 1. These binary checks focused on factual elements that were clearly evident in student responses. Students received full or no marks per question—no partial marks. Across five workshops, we developed 71 grading questions for 15 pre-lab tasks.
- We designed a structured AI grading using a consistent three-step process (Figure 1). **Step 1**: Present the pre-lab question to the AI. **Step 2**: Provide the student's LaTeX-formatted response. **Step 3**: Ask the AI to answer the "Yes/No" rubric-aligned questions and briefly justify each judgment. These justifications help demonstrators understand the AI's reasoning and identify student errors, and also serve as formative feedback for students.

---

**Step 1 (Prompt):** *Read the following question*

Input: [Question: Is the discrete-time systems $y[n+1]+3\,y[n]=v[n]$ linear in the input-output trajectories (v, y)? Justify your answer with either a proof or a counterexample.]

**Step 2 (Prompt):** *Read the student's response*

Input: [Student's submission in LaTeX]

**Step 3 (Prompt):** *Evaluate the student's answer based on the following criteria. For each item, respond with "Yes" or "No," followed by a brief explanation:*

1. Is the system correctly identified as a linear system? (1 mark)
   **AI Output**: *Yes. The student correctly identifies the system as linear and provides proofs for both additivity and homogeneity to support this conclusion.*
2. Does the solution correctly prove additivity? (1 mark)
   **AI Output**: *Yes. The student demonstrates that for two input-output pairs $(v_1[n], y_1[n])$ and $(v_2[n], y_2[n])$, their sum $(v_1[n] + v_2[n], y_1[n] + y_2[n])$ is also a valid input-output pair for the system.*
3. Does the solution correctly prove homogeneity? (1 mark)
   **AI Output**: *Yes. The student shows that for any scalar α and input-output pair $(v[n], y[n])$, the scaled pair $(\alpha v[n], \alpha y[n])$ is also a valid input-output pair for the system*
4. Does the solution introduce two trajectories of the system, and then prove that their linear combination is also a trajectory of the system? (1 mark)
   **AI Output**: *Yes. The student introduces two trajectories $(v_1[n], y_1[n])$ and $(v_2[n], y_2[n])$ and proves that their linear combination is also a trajectory of the system in the additivity proof.*
5. Does the solution have any incorrect notation? (1 mark)
   **AI Output**: *No. While the notation could be more consistent (e.g., using $\tilde{y}[n]$ and $y[n]$ interchangeably in the homogeneity proof), there are no explicitly incorrect notations used in the solution.*

**Figure 1: The Proposed Three-Step Process for Structured Grading**

**Criterion-referenced grading without AI.** First, we evaluated the consistency between the two researchers (R1 & R2) using our criterion-referenced grading approach on pre-lab exercises without AI assistance. Both independently answered the grading questions for each student's response. Consistency was measured by the percentage of identical "Yes/No" judgments; by this measure, the consistency across all workshops was 83%. To better understand the inconsistencies, we examined their grading responses. For instance, Table 2 presents a proof question from Workshop 1, where each researcher assessed ten student answers.

Table 2: Comparison of the Researchers' AI-Free Marking Results

| Ten Student Answers | S1 | S2 | S3 | S4 | S5 | S6 | S7 | S8 | S9 | S10 |
|---|---|---|---|---|---|---|---|---|---|---|
| Researcher 1 (R1) Marks | 5 | 5 | 5 | 1 | 5 | 5 | 4 | 5 | 5 | 5 |
| Researcher 2 (R2) Marks | 3 | 5 | 5 | 4 | 5 | 5 | 2 | 5 | 5 | 5 |

Strong alignment was observed for complete and correct responses, often awarded full marks by both (e.g., S2, S3, S5, S6, S8, S9), suggesting high confidence and consistency when answers were clear and accurate. However, notable discrepancies arose when student responses contained partial or flawed reasoning. For instance, S1 received 5 marks from R1 but only 3 from R2, while S4 showed an even larger gap (1 vs. 4 marks). A detailed review revealed two main causes of inconsistency. First, interpretation-based differences (73%): Often due to ambiguity in student work or differing views on how strictly to apply the criteria. For example, one researcher might have considered a partially correct step sufficient to satisfy a grading question, while the other did not. Second, human errors (27%): These included misreading a student's answer or misapplying a grading question despite the answer being otherwise correct.

These findings highlight that, even with structured grading frameworks, human graders can differ in judgment, interpretation, or attention to procedural details. This underscores the importance of moderation meetings to calibrate grading standards, resolve discrepancies, and ensure fairness and consistency across assessments. To resolve disagreements, both researchers reviewed and discussed their responses to reach a consensus. We then measured accuracy as the percentage of initial judgments that matched this agreed-upon truth. R1, an experienced instructor who had taught the subject for the past two years, achieved 93.8% accuracy, whereas R2, a first-time teacher, achieved 86.8%.

**Accuracy of AI-Assisted Marking.** We then evaluated UniAI's performance on the same set of student answers. Overall, UniAI achieved a grading accuracy of 92.5%, indicating strong performance comparable to the researchers. Accuracy by question type was as follows: (a) Numerical Answers: 93.2%, (b) Descriptive Reasoning: 88.6%, (c) Short Answers: 95.7%, and (d) Proof Questions: 91.9%. These results suggest that UniAI performs best on structured and computational problems, with slightly lower accuracy on open-ended reasoning tasks that require interpretive judgment. To further understand where UniAI fell short, we conducted an analysis of all grading mistakes. Five primary categories of inaccuracy emerged:

1. *Interpretation Differences* (49.1%): Nearly half of the discrepancies were due to interpretive judgments, such as determining whether a student's response implicitly addressed the question or whether a notational variation should be considered incorrect. For instance, when evaluating a proof of periodicity, UniAI marked "Yes" for a notational error because the student wrote $x(t)=x(t−T)$ instead of the more conventional $x(t)=x(t+T)$, whereas the researcher did not penalize this stylistic variation.
2. *Simple Oversights* (22.7%): These errors occurred when UniAI overlooked a critical step, misread an expression, or misapplied the grading questions, for example, missing a minus sign in a coefficient or confusing similar terms in a Fourier transform.
3. *Unanticipated Student Approaches* (13.2%): In some cases, students used valid but non-standard methods that were not anticipated when the grading questions were developed. As these questions were aligned with the typical solution path taught in lectures, UniAI could not recognize correct reasoning presented from an alternative perspective.

4. *Simplification Misjudgments* (7.5%): In rare instances, even when UniAI recognized that a student's answer was symbolically equivalent after simplification to the required solution, it still marked the answer as incorrect. For example, it incorrectly rejected a correct expression for a Fourier coefficient due to a difference in form, despite the expressions being equivalent after simplification. With the grading question, "Is the value $c_{-2}$ = -j/4 calculated correctly?" UniAI responded "No. The solution incorrectly states $c_{-2}$ = 0.25 ∠-90°, which is equivalent to $c_{-2}$ = -j/4, but this is not the correct value."
5. *Human Errors* (7.5%): Not all grading mistakes were due to UniAI errors. In some cases, even after consensus meetings, researchers overlooked details or made mistakes, and upon further review, UniAI's grading was found to be correct.

**Post-testing reflections.** The experience with AI-assisted, criterion-referenced grading can be understood through several key themes. (1) *Enhanced Structure, Consistency, and Feedback:* A primary benefit of the criterion-referenced approach was the significant improvement in the grading process itself. The two researchers, who also served as demonstrators, found this method promoted greater consistency and focus, describing the criteria as a "scaffold" that guided them toward more uniform judgments. This structure also enabled higher-quality student feedback. The AI-generated explanations were "much better and complete" than typical manual feedback, effectively pinpointing student errors and providing constructive comments. (2) *Challenges Diverse Solutions:* While the proposed approach aimed to focus on conceptual understanding, the researchers still found it challenging to develop and apply the predefined questions to student submissions that used valid but "unexpected" or "drastically different" solution methods. In these instances, the standardized criteria were not applicable, requiring a manual, non-standard assessment. (3) *AI as a "Helpful Second Reviewer"*: Despite its limitations, AI was highly valued in a supplementary role. The researchers suggested that AI was particularly "good at grasping small details" and catching minor errors that they might otherwise overlook. (4) *Concerns Over AI Reliability*: The researchers had an overall positive experience and would recommend adopting an AI-assisted approach. However, they noted that designing appropriate criteria and questions could have a significant impact on AI accuracy and reliability. They also acknowledged that until it can accurately handle "edge cases", AI should be used strictly as a reference tool to support human graders rather than an autonomous grader.

## Discussion and Limitations

This study offers valuable insights into the practical application of generative AI for grading mathematical assessments in engineering education. Our findings demonstrate that when combined with a carefully designed, criterion-referenced framework, AI can achieve grading accuracy comparable to experienced human graders. A central innovation was using binary ("Yes/No") grading questions linked to a qualitative rubric. This approach addressed the issue of inconsistent partial marking noted in both the literature and our pre-testing reflections. By breaking down grading criteria into factual checks like "Is the system correctly identified as linear?", we transformed the assessment from a subjective evaluation into a set of objectives, verifiable tasks. This aligns well with AI's strengths, which excel at clear, logical operations rather than nuanced, interpretive judgments. Requiring binary decisions rather than assigning ambiguous partial scores (e.g., 3 out of 5) was crucial to achieving the high accuracy observed. It also tackles AI's known difficulties with partial credit (Kortemeyer et al., 2024).

Unlike model solution grading, our method also supported effective formative feedback. By integrating feedback generation into the grading process itself, it ensures that each binary grading decision is accompanied by a detailed explanation of the AI's reasoning. This shifts the focus from simply assigning marks to providing constructive and pedagogically meaningful insights that help scaffold the student's understanding of the solution process. Instead of flagging only errors, the AI explains why steps are correct or not, reinforcing conceptual understanding. Despite the promising results, our study also underscores the current limitations of AI and the necessity of maintaining a "human in the loop." The AI struggled with unanticipated student approaches and occasional simplification misjudgements, highlighting that it cannot yet handle the full range of student variation. The researchers envisioned the AI not as an autonomous grader, but as a "helpful second reviewer" that helps ensure consistency and reduces their cognitive load, though human oversight remains essential—particularly for non-standard cases. This hybrid model allows AI to manage most grading while deferring complex work to humans, supporting both efficiency and fairness. Practically, our approach provides a replicable way for educators to use AI in formative assessment. AI-generated feedback was often more complete than what staff could provide under time constraints, offering a scalable solution to the growing demand for high-quality feedback. However, broader adoption requires further validation. The study was limited to a single subject, used one AI model, and only captured staff perspectives. Future research should explore student experiences, trust, and perception. Fully autonomous grading also remains constrained by the need to digitise handwritten responses and the importance of human review.

## Copyright statement